\documentclass[aps,prl,superscriptaddress,showpacs,twocolumn]{revtex4}
\usepackage{amsmath,amsfonts,bm,graphicx,amssymb}
\newcommand{\ket}[1]{|\mbox{$#1$}\rangle}
\begin{document}

\title{Spin-Photon Entangling Diode}
\author{Christian Flindt}
\affiliation{MIC -- Department of Micro and Nanotechnology, Technical University of Denmark, Kongens Lyngby
2800, Denmark} \affiliation{QUANTOP and the Niels Bohr institute, University of Copenhagen, Copenhagen
2100, Denmark} \affiliation{Department of Physics, Harvard University, Massachusetts 02138, USA}
\author{Anders S. S\o rensen}
\affiliation{QUANTOP and the Niels Bohr institute, University of Copenhagen, Copenhagen
2100, Denmark}
\author{Mikhail D. Lukin}
\affiliation{Department of Physics, Harvard University, Massachusetts 02138, USA}
\author{Jacob M. Taylor}
\affiliation{Department of Physics, Harvard University,
Massachusetts 02138, USA} \affiliation{Department of Physics,
Massachusetts Institute of Technology, Massachusetts 02139, USA}
\date{\today}
\begin{abstract}
  We propose a semiconductor device that
  can electrically generate entangled electron spin-photon states, providing a building block for
  entanglement  of distant spins. The device consists of a \emph{p-i-n} diode structure that incorporates a coupled double quantum dot.
We show that electronic control of the diode
  bias and local gating allow for the generation of
  single photons that are entangled with a robust quantum memory
  based on the electron spins. Practical performance of this approach  to controlled spin-photon entanglement is analyzed.
\end{abstract}



\maketitle

Many practical approaches to quantum communication and computation
rely upon interfacing stable quantum systems, which provide a good
quantum memory, with carriers of quantum information (optical
photons) at the level of single quanta. One promising approach to
quantum memory  uses electron spins confined in semiconductor
quantum dots.  Quantum dots in diode structures can also be used
for creating devices with novel electronic and optical properties.
In particular, the Coulomb blockade exhibited by quantum dots is
being used in experiments involving single charge and spin
transport and manipulation \cite{kroutvar04,Petta:2005,koppens06}
as well as for optical experiments such as generation of
single-photons \cite{Imamoglu:1994,kim99,michler00,Bennett:2006}.
Application of these systems for  realization of quantum
communication and computation protocols  is a vibrant area of
research~\cite{Chen:00,Gywat:02,waks02,Badolato:2005,Krenner:2006,Stinaff:2006,Stevenson:06}.

In this Letter we propose and analyze a novel semiconductor device
 in which an electrically pumped diode structure can combine controlled photonic interface with stable quantum memory. Such a device features purely electrical control over photonic and spin degrees of freedom. Specifically we show that it can be used for controlled generation of  entangled states between the frequency of an outgoing photon and the spin state of the 
 electrons in a double quantum dot
in the insulating layer of the diode similar to recent laser
driven experiments in atomic systems
\cite{Blinov:2004,Volz:2006,Rosenfeld:2007}. Using recently
demonstrated techniques \cite{taylor05}, the double-dot spin
states can provide a robust quantum memory for long-term
information storage, while outgoing photons can be used for
probabilistic generation of long-distance entanglement in direct
analogy to approaches being explored for atomic
systems~\cite{Duan:2001}. Finally, when integrated with
gate-controlled quantum dot systems~\cite {Engel:2006}, this
device could also form a building block for scalable quantum
computation.

\begin{figure}[ht!]
\begin{center}
\includegraphics[width=0.48\textwidth, trim = 0 0 0 0, clip]{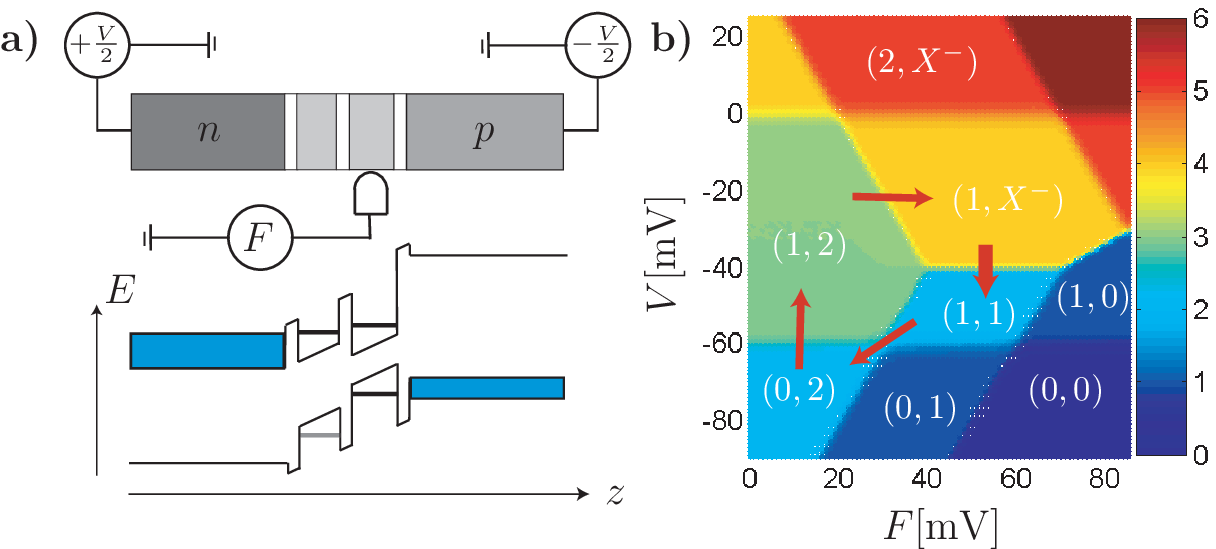}
\includegraphics[width=0.40\textwidth, trim = 0 0 0 0, clip]{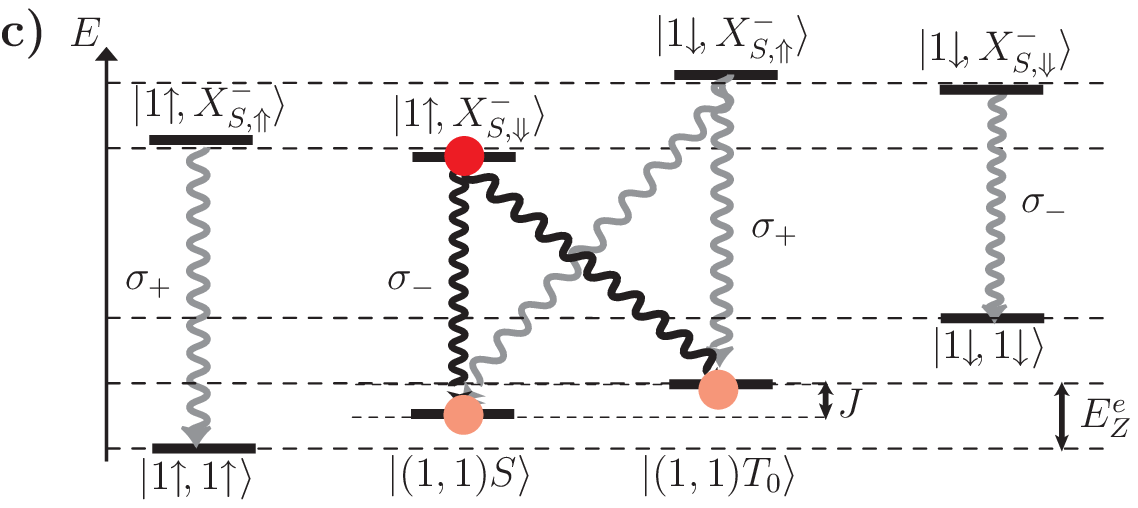}
\end{center}
\caption{(color online). Diode structure,
  charge stability diagram, and decay paths. a) The device consists of a double quantum
  dot within the intrinsic region of a \mbox{\emph{p-i-n}} diode
  structure. A schematic band-edge diagram is shown below. The left hole state (gray) is assumed to be
  energetically out of reach.
  b) The stable charge configuration of the double quantum dot as
  function of the bias $V$ and the local gate $F$.
  Colors denote the total number of charges, while the labeling $(n,m)$ corresponds to the number of
   electrons on the left ($n$) and right ($m$) dot, replaced by $X^-$ for dots containing a negatively charged exciton consisting of two electrons and a hole.
   As discussed in the text, the charging sequence indicated by arrows preferentially initializes the dots in the state $|1\!\!\uparrow,X^-_{S,\Downarrow}\rangle$.
   c) Initial and final states for excitonic recombination together with the polarization of the emitted photon. The desired decay processes  (shown in black) are selected by filtering $\sigma_-$ polarization.}
\label{fig:device}
\end{figure}

The basic idea of our approach can be understood by considering the
semiconductor nanowire shown in Fig.\ \ref{fig:device}a, in which a
Coulomb-blockade double quantum dot is sandwiched between the
positively and negatively doped semiconductor regions, forming a \mbox{\emph{p-i-n}} diode. By
manipulating the bias across the diode and the
local gate, we can control the injection of electrons and holes
into the double dot at the level of single charges. This allows us
to electrically prepare a metastable exciton complex that decays by
electron-hole recombination to a charge configuration with a single
electron in each of the two dots. When the two-electron spin states
are split by the exchange coupling, the left circularly polarized
photon that is emitted under the electron-hole recombination process
will be frequency-entangled with the spin state of
the remaining electrons (Fig.~\ref{fig:device}c).

Double quantum dots can be grown inside \mbox{\emph{p-i-n}}
junctions with techniques similar to those recently used to
fabricate single quantum dot nanowire LEDs
\cite{Borgstrom:2005,Minot:2007}. Alternatively self-assembled
dots on wafers can be used~\cite{Solomon:1996}. As illustrated in
Fig.\ \ref{fig:device}a the chemical potentials of the $p$ and $n$
regions on each side of the central (intrinsic) region can be
controlled by applying a bias across the device, while a gate
electrode nearby the double dot can be used to tune the levels
inside the dots independently of the applied bias.  In what
follows we focus on \mbox{III-V} semiconductors, but in practice
other optically active materials can be used.

The electrostatic properties  of the device can be visualized by using a charge stability diagram
\cite{vanderwiel:2003}. We describe the charge degrees of freedom of the double quantum dot using the Hamiltonian
\begin{equation}
\label{eq:DQDmodel}
\hat{H}_{DQD}=\hat{H}_{ee}^l+\hat{H}_{ee}^r+\hat{H}_{hh}^r+\hat{H}_{eh}^r+\hat{H}_{\tau}+\hat{H}_F,
\end{equation}
where the Coulomb repulsion between similar charges $(q=e,h)$ on the left or right dot reads
$\hat{H}^s_{qq}=U_{qq}\hat{n}_q^s(\hat{n}_q^s-1)/2$, $s=l,r$, while the Coulomb attraction between electrons
and hole in the right dot reads $\hat{H}_{eh}^r=-U_{eh}\hat{n}_e^r\hat{n}_h^r$. Here $\hat{n}_{e(h)}^{l\{r\}}$
is the operator for the number of electrons (holes) in the left \{right\} dot. Tunneling between the dots with
tunnel coupling $\tau$ is contained in the term $\hat{H}_{\tau}$, while
$\hat{H}_F=-eF(\hat{n}_h^r-\hat{n}_e^r)$ incorporates the shift of the electron and hole states in the right
 dot due to the local gate $F$. Electrons with spin $\sigma=\pm 1/2$ on the left and right dots are created by
$\hat{c}_{l,\sigma}^{\dag}$ and $\hat{c}_{r,\sigma}^{\dag}$, respectively, while heavy holes in the right dot
are described by $\hat{d}_{r,\Sigma}^{\dag}$, with $\Sigma = \pm 3/2$. The hole states of the left dot are
assumed to remain unoccupied due to band-gap differences between the two dots, which can be induced, e.g., by  strain.
Electrons are injected into the left dot from the electron Fermi sea in the $n$-region at chemical potential
$\mu_n$, while holes are injected into the right dot from the hole Fermi sea in the $p$-region at chemical
potential $\mu_p$. Assuming weak coupling to the electron and hole Fermi seas, we solve numerically the master equation for the probability of occupying the different many-body eigenstates of the double dot and find
the stable charge configurations for the chosen parameter range.

In Fig.\ \ref{fig:device}b we show the resulting charge stability
diagram, where the total number of charges on the double quantum
dot is given as function of the bias across the device $V$ and the
local gate $F$. Here, $\mu_n=\mu_n^0+eV/2$ and
$\mu_p=\mu_p^0-eV/2$, where $\mu^0_{n}$ and $\mu_0^{p}$, given by
the doping levels of the $n$ and $p$ regions, respectively,
determine the filling of the double dot without applied voltages.
In the numerical calculations, the values of $\mu_n^0$ and
$\mu_h^0$ were used to fix the occupations of the dots at $V,F=0$.
For the shown charge stability diagram we have used $U_{ee}\simeq
30$ meV, $U_{hh}\simeq 50$ meV, $U_{eh}\simeq 40$ meV, $\tau=1$
meV, and the tunnel couplings to the electron and hole Fermi seas
being identical and much smaller than the temperature $T=4$ K.
With a given setting of $V$ and $F$ the system rapidly reaches the
corresponding stable charge configuration. On the figure we also
indicate the charge configuration of each of the two dots, where
the labeling $(n,m)$ refers to the charges on the left dot ($n$)
and the right dot ($m$), respectively. For configurations with no
holes, the two labels correspond to the number of electrons in the
left and right dot, respectively, while the symbol $X^{-}$ denotes
a negatively charged exciton consisting of two electrons and a
hole. Such excitonic states have previously successfully been
generated and controlled optically \cite{Dutt:2005, Atature:2006},
but the procedure presented here works all electrically and
thereby does not require any laser control. An external magnetic
field $B$ is applied to the system parallel to the light emission
and growth axis (the $z$-axis on Fig.\ \ref{fig:device}a), i.e.,
in a Faraday configuration. In order to have reliable electron
spin state preparation, we will require the Fermi seas to be
sufficiently cold: $k_B T \ll |g^* \mu_B B|$ (see below).

We now describe the charge injection sequence indicated by arrows
in Fig.\ \ref{fig:device}b. The sequence allows us repeatedly to
prepare a desired spin and charge configuration, by injecting
charges one at a time, and direct its decay and corresponding
photon emission process. We assume that single charges may be
injected faster than the spontaneous decay time ($\sim 1$ ns for
GaAs self-assembled quantum dots). By controlling the bias $V$ and
the local gate $F$, the system is first put in the charge
configuration $(0,2)$. The expected ground state spin
configuration of this state is a singlet due to the tight
confinement of the two electrons to a single dot: $|0,2S\rangle =
\hat{c}_{r,\uparrow}^{\dag} \hat{c}_{r,\downarrow}^{\dag}
\ket{0}$. After preparing the $|0,2S\rangle$ state, an additional
spin-up electron is added to the left dot by increasing the bias,
taking $|0,2S\rangle$ to $|1\uparrow,2S\rangle =
\hat{c}_{l,\uparrow}^{\dag} \hat{c}_{r,\uparrow}^{\dag}
\hat{c}_{r,\downarrow}^{\dag} \ket{0}$. A heavy hole with spin
$\Sigma = \Downarrow$ is now added to the right dot by control of
the local gate, yielding the state $
|1\uparrow,X^{-}_{S,\Downarrow}\rangle =
\hat{d}_{r,\Downarrow}^{\dag}
\hat{c}_{l,\uparrow}^{\dag}\hat{c}_{r,\uparrow}^{\dag}
\hat{c}_{r,\downarrow}^{\dag} \ket{0}$, which we expect to decay
to $|1\uparrow,1\downarrow\rangle =
\hat{c}_{l,\uparrow}^{\dag}\hat{c}_{r,\downarrow}^{\dag} \ket{0}$
via excitonic recombination ($\hat{d}_{r,\Downarrow}
\hat{c}_{r,\uparrow}$). However, before recombination takes place
we rapidly (i.e., faster than the decay rate) move to the region,
where $(1,1)$ is the stable charge configuration, hereby
preventing emission of more than a single photon (by re-filling of
an electron and a hole) in each cycle of the sequence.

The exciton decay couples the state
$|1\uparrow,X^{-}_{S,\Downarrow}\rangle$ to
\mbox{$|1\uparrow,1\downarrow\rangle$}. With finite tunnel
coupling between the left and right dots, this may be written as a
superposition
$|1\uparrow,1\downarrow\rangle=(|(1,1)S\rangle+|(1,1)T_0\rangle)/\sqrt{2}$,
of the exchange-split singlet and triplet eigenstates
$|(1,1)S(T_0)\rangle=2^{-1/2}(\hat{c}_{l,\uparrow}^{\dag}\hat{c}_{r,\downarrow}^{\dag}\pm\hat{c}_{l,\downarrow}^{\dag}\hat{c}_{r,\uparrow}^{\dag})\ket{0}$.
Since $S$ and $T_0$ have different energies, the frequency of the
outgoing photon will be entangled with the spin state (see Fig.\
\ref{fig:device}c and Eq. (\ref{eq_state}) below). These $S$-$T_0$
spin states were used in recent double dot experiments where it
was shown that they form a decoherence free subspace when
manipulated with fast spin-echo pulses
\cite{Petta:2005,Levy:2002,Taylor:2005}. With the system in
Faraday configuration the spin of the hole determines the
polarization of the emitted photon. While a spin-$\Downarrow$
heavy-hole recombines with a spin-$\uparrow$ electron under
emission of a left-hand circularly polarized ($\sigma_{-}$)
photon, a spin-$\Uparrow$ heavy-hole recombines with a
spin-$\downarrow$ electron under emission of a right-hand
circularly polarized ($\sigma_{+}$) photon. By suitable
polarization filtering it is thus possible to exclude photons that
have been emitted with the heavy-hole incorrectly being in the
spin-$\Uparrow$ state.

\begin{figure}
\includegraphics[width=0.46\textwidth, trim = 0 0 0 0, clip]{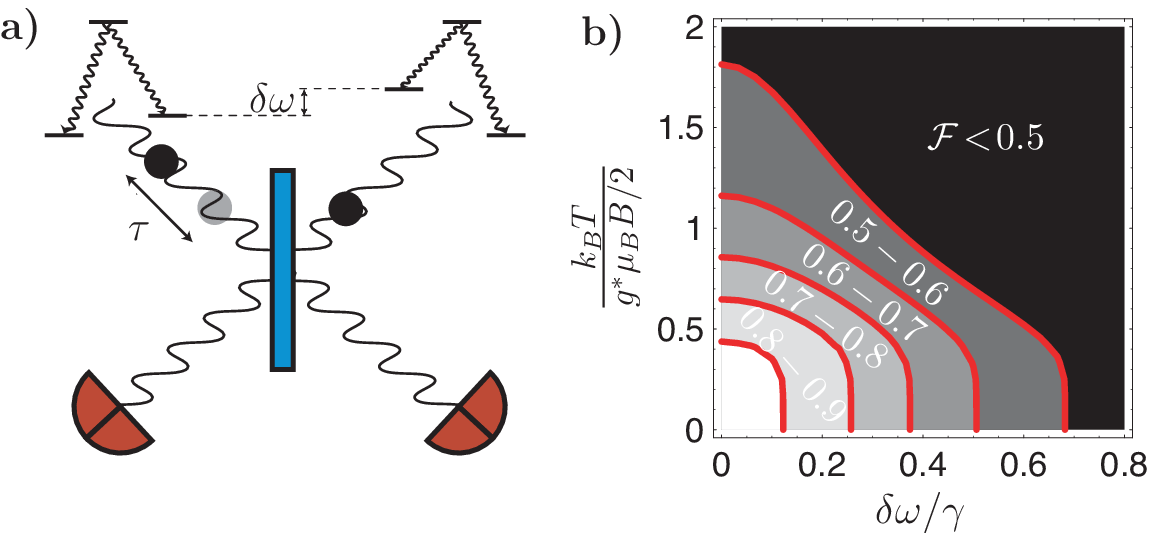}
 \caption{Beam splitter setup and entanglement fidelity. a) Photon interference leading to entanglement between two devices.
The entanglement fidelity may be reduced due to different arrival
times $\tau$ at the beam splitter, mismatch $\delta\omega$ between
energy splittings in the two devices, and incorrect spin
initialization (not indicated). b) Entanglement fidelity
$\mathcal{F}$ as function of temperature $T$
 and energy mismatch $\delta\omega$. For the calculations we have used $\gamma\times\bar{\tau}=0.03$. In the white region $\mathcal{F}>0.9$.}
 \label{fig:beam_splitter}
\end{figure}

The resulting spin-photon entangled state can be used for
generating spin-spin entanglement between two remote devices by
interfering the emitted photons on a beam splitter as shown in
Fig.\ \ref{fig:beam_splitter}a \cite{Duan:2006}.
 If the spin state in both devices are identical,
both incoming photons can be  mode matched in space, frequency,
and time, so that Hong-Ou-Mandel bunching will occur, leading to
photon detection in only one arm of the beam splitter.  On the
other hand, if the spin states are different, the photons are
distinguishable, and no ``bunching'' will occur. A photon detection
in each arm of the beam splitter therefore leads to an entangled
state of the spins in the spatially separated devices
$|\Psi_{12}\rangle=(|S\rangle_1|T_0\rangle_2-|T_0\rangle_1|S\rangle_2)/\sqrt{2}$,
where we have omitted the charge labeling $(1,1)$. In the
following we consider the distinguishability of our outgoing
photons to determine the requirements for such entanglement
generation.

We first consider spontaneous decay associated with electron-hole
recombination in a single device. The process can be described
within the framework of Wigner-Weisskopf theory yielding a
characteristic decay rate $\gamma$. We note that the ground state
charge configuration $(1,1)$ may also be reached by the
electron-hole pair tunneling back into the Fermi seas with rate
$\Gamma_{\! o}$ rather than recombining. This does not impact the
fidelity of entanglement but does reduce the success probability.
After spontaneous decay has taken place the combined state
$|\Phi\rangle$ (conditioned on electron-hole recombination) of the
charges and the photon field reads
\begin{equation}
|\Phi\rangle =
\frac{1}{\sqrt{2}}\left[|S\rangle\otimes\hat{\xi}^{\dagger}(\omega_S)|0\rangle+|T_0\rangle\otimes\hat{\xi}^{\dagger}(\omega_{T_0})|0\rangle\right],
\label{eq_state}
\end{equation}
where
$\hat{\xi}^{\dagger}(\omega)=\sum_{k}\xi(\omega,k)\hat{a}^{\dagger}_k$
with
$\xi(\omega,k)=\frac{1}{\sqrt{2\pi}}\frac{\sqrt{\gamma}e^{-ikz_0}}{(\omega_k-\omega)+i\gamma/2}$
and $\hat{a}^{\dagger}_k$ being the creation operator for photons of
mode $k$. Here, $|0\rangle$ is the vacuum state of the photon field,
while the position of the double quantum dot is $z_0$, and
$\omega_S$ and $\omega_{T_0}$ denote the splittings between the
excited state and the singlet and triplet groundstates,
respectively, so that $|\omega_S-\omega_{T_0}|$ equals the exchange
coupling $J$. The width of the photon wavepacket is given by
$\gamma=\gamma_S+\gamma_T+\Gamma_{\! o}$, and above we have taken
the same rates for the two decay paths, $\gamma_S=\gamma_T$,
resulting in equal branching ratios for the two processes.

We now consider to the beam-splitter setup depicted in Fig.\
\ref{fig:beam_splitter}a and consider two photons emitted by
similar devices. With probability $1/2$ the two photons are in
states corresponding to the same spin state of the electrons in
the two devices (both singlet or both triplet) and with
probability $1/2$ in states corresponding to different spin
states. The probability of detecting two photons, in states
$|\Psi_L\rangle=\xi^{\dagger}_L(\omega_L)|0\rangle$ and
$|\Psi_R\rangle=\xi^{\dagger}_R(\omega_R)|0\rangle$, respectively,
at different detectors (denoted $L$ and $R$) after they have
scattered on the 50/50 beam splitter is $P(1_L,1_R) =
(1-|\mathcal{J}|^2)/2$, where
$\mathcal{J}\equiv\sum_{k}\xi_L(\omega_L,k)\xi^*_R(\omega_R,k)$ is
the overlap of the wavepacket amplitudes. With $\omega_L=\omega_S$
and  $\omega_R=\omega_{T_0}$, we find
\begin{equation}
P(1_L,1_R) = \frac{1}{2}\left(1-\frac{\gamma^2}{\gamma^2+J^2}\right).
\label{eq_exchange}
\end{equation}
Typical electron-hole recombination rates are on the order of GHz,
and the width of the wavepackets $\gamma$ therefore on the order
of $\mu$eVs, which is typically smaller than the exchange coupling
$J$ between electrons in tunnel-coupled quantum dots, which can
reach values on the order of meVs. Thus we expect $P(1_L,1_R) \sim
1/2$ and a corresponding success rate of $\eta^2/4$ for detecting
the two photons at different detectors, where $\eta$ is the
combined single photon emission and detection probability.

Besides the success rate we need to consider the fidelity
$\mathcal{F}= \langle\Psi_{12}|\hat{\rho}_s|\Psi_{12}\rangle$ of
the entangling procedure, where $|\Psi_{12}\rangle$ is the desired
state, and $\hat{\rho}_s$ is the reduced density matrix for the
spins in the two devices. The possible error processes include
wrong initialization of the spin states, jitter in the hole
injection and path length differences leading to different arrival
times at the beam splitter, and different energy splittings
between the excited state and the ground states in the two
devices. To evaluate the effect of jitter and the different energy
splittings we write the full density matrix of the spins and
photons as
$\hat{\rho}=|\Phi\rangle_1|\widetilde{\Phi}\rangle_2{}_2\langle\widetilde{\Phi}|_1\langle\Phi|$
where both $|\Phi\rangle_1$ and $|\widetilde{\Phi}\rangle_2$ are
of the form given in Eq.\ (\ref{eq_state}), but
$|\widetilde{\Phi}\rangle_2$ has a component on a field mode
 perpendicular to the field mode
emitted by device number 1, i.e., for
$|\widetilde{\Phi}\rangle_2$, $\hat{\xi}^{\dagger}$ is replaced by
$\mathcal{J}\hat{\xi}^{\dagger}+\sqrt{1-|\mathcal{J}|^2}\hat{\xi}_
{err}^{\dagger}$. Here, $\hat{\xi}_ {err}^{\dagger}$ creates a
photon in an undesired mode due to the jitter and energy shifts,
and $\mathcal{J}$ denotes the corresponding wavepacket overlap.

Conditioned on clicks in different detectors after the beam
splitter, we find that the erroneous field component generates the
spin states $|T_0\rangle_1|T_0\rangle_2$,
$|S\rangle_1|S\rangle_2$, $|T_0\rangle_1|S\rangle_2$, and
$|S\rangle_1|T_0\rangle_2$  with equal probability
$(1-|\mathcal{J}|^2)/(4-3|\mathcal{J}|^2)$.
 The fidelities corresponding to
these states are 0, 0, 1/2, and 1/2, respectively. The desired
state
$(|S\rangle_1|T_0\rangle_2-|T_0\rangle_1|S\rangle_2)/\sqrt{2}$
(with fidelity 1) is generated with probability
$|\mathcal{J}|^2/(4-3|\mathcal{J}|^2)$. Combining these numbers,
we find the fidelity $1/(4-3|\mathcal{J}|^2)$, which, however,
does not yet include the possibility of wrong spin initialization.

In thermal equilibrium the probability of initializing the wrong
spin state $|\!\downarrow\rangle$ in the left dot is given by the
Boltzmann factor $p_{\downarrow}\propto e^{-g^*\mu_BB/2k_BT}$.
Wrong initialization of the spin in one or both of the devices
leads to generation of states with fidelity 0. The probability of
detection at different detectors due to a wrong spin in one or
both of the devices is bounded from above by
$\eta^2(2p_{\downarrow}-p^2_{\downarrow})/2$, which we use in the
following. Including this estimate for the effect of wrong spin
initialization in the above expression for the fidelity, we find
\begin{equation}
\mathcal{F}=\frac{1}{4-3|\mathcal{J}|^2}\times\frac{1}{1+2(2p_{\downarrow}-p^2_{\downarrow})}.
\label{eq_fidelity}
\end{equation}
Two photons created by $\hat{\xi}^{\dagger}(\omega)$ and
$\hat{\xi}^{\dagger}(\omega+\delta\omega)$ with a time difference
$\tau$ have the wavepacket overlap
$|\mathcal{J}|^2=\frac{\gamma^2e^{-\gamma|\tau|}}{\delta\omega^2+\gamma^2}$.
For the time difference, we assume a Gaussian probability
distribution with width $\bar{\tau}$,
 $\mathcal{P}(\tau)\propto
e^{-(\tau/\bar{\tau})^2/2}$. This distribution is relevant when
noise in the gates controlling the hole injection is responsible
for the photons being created at different times or when the
optical paths do not have exactly the same length. When evaluating
the fidelity we average the expression in Eq.\ (\ref{eq_fidelity})
with respect to the Gaussian distribution.

In Fig.\ \ref{fig:beam_splitter}b we show the fidelity as function
of temperature and energy mismatch. We see that with realistic
parameters it is possible to obtain a high degree of fidelity,
$\mathcal{F}>0.9$, and even with temperatures comparable to the
Zeeman energy, the fidelity may be larger than 0.5, the lower
bound for the use of entanglement purification protocols
\cite{Briegel:1998}. Furthermore, the loss of fidelity due to time
jitter or energy mismatch may be suppressed by gating the
detectors in time, thereby improving the shown results.

In  conclusion, we have presented a proposal for an
all-electrically controlled device for long-range electron-spin
entanglement and shown that entanglement can be generated with a
high degree of fidelity using available experimental techniques.
When combined with existing quantum optical methods and
solid-state technologies for electron spin manipulation and
detection, our proposed device may form an important building
block in future quantum communication and information processing
architectures.

We thank M.\ G.\ Dutt, K.\ Flensberg, A.\ Imamoglu, and M.\ Stopa
for helpful conversations. The work was supported by the
Denmark-America Foundation, Pappalardo Fellowship, the Danish
Natural Science Research Council, NSF, DTO, and the Packard
foundation.


\end{document}